\documentclass[aps,prl,twocolumn,superscriptaddress,showpacs,letterpaper]{revtex4}

\usepackage{graphicx}
\usepackage{amsmath, amsthm, amssymb}

\begin{document}

\title{Observation of Orbitally Excited $B_s$ Mesons}

\affiliation{Institute of Physics, Academia Sinica, Taipei, Taiwan 11529, Republic of China} 
\affiliation{Argonne National Laboratory, Argonne, Illinois 60439} 
\affiliation{Institut de Fisica d'Altes Energies, Universitat Autonoma de Barcelona, E-08193, Bellaterra (Barcelona), Spain} 
\affiliation{Baylor University, Waco, Texas  76798} 
\affiliation{Istituto Nazionale di Fisica Nucleare, University of Bologna, I-40127 Bologna, Italy} 
\affiliation{Brandeis University, Waltham, Massachusetts 02254} 
\affiliation{University of California, Davis, Davis, California  95616} 
\affiliation{University of California, Los Angeles, Los Angeles, California  90024} 
\affiliation{University of California, San Diego, La Jolla, California  92093} 
\affiliation{University of California, Santa Barbara, Santa Barbara, California 93106} 
\affiliation{Instituto de Fisica de Cantabria, CSIC-University of Cantabria, 39005 Santander, Spain} 
\affiliation{Carnegie Mellon University, Pittsburgh, PA  15213} 
\affiliation{Enrico Fermi Institute, University of Chicago, Chicago, Illinois 60637} 
\affiliation{Comenius University, 842 48 Bratislava, Slovakia; Institute of Experimental Physics, 040 01 Kosice, Slovakia} 
\affiliation{Joint Institute for Nuclear Research, RU-141980 Dubna, Russia} 
\affiliation{Duke University, Durham, North Carolina  27708} 
\affiliation{Fermi National Accelerator Laboratory, Batavia, Illinois 60510} 
\affiliation{University of Florida, Gainesville, Florida  32611} 
\affiliation{Laboratori Nazionali di Frascati, Istituto Nazionale di Fisica Nucleare, I-00044 Frascati, Italy} 
\affiliation{University of Geneva, CH-1211 Geneva 4, Switzerland} 
\affiliation{Glasgow University, Glasgow G12 8QQ, United Kingdom} 
\affiliation{Harvard University, Cambridge, Massachusetts 02138} 
\affiliation{Division of High Energy Physics, Department of Physics, University of Helsinki and Helsinki Institute of Physics, FIN-00014, Helsinki, Finland} 
\affiliation{University of Illinois, Urbana, Illinois 61801} 
\affiliation{The Johns Hopkins University, Baltimore, Maryland 21218} 
\affiliation{Institut f\"{u}r Experimentelle Kernphysik, Universit\"{a}t Karlsruhe, 76128 Karlsruhe, Germany} 
\affiliation{Center for High Energy Physics: Kyungpook National University, Daegu 702-701, Korea; Seoul National University, Seoul 151-742, Korea; SungKyunKwan University, Suwon 440-746, Korea; Korea Institute of Science and Technology Information, Daejeon, 305-806, Korea; Chonnam National University, Gwangju, 500-757, Korea} 
\affiliation{Ernest Orlando Lawrence Berkeley National Laboratory, Berkeley, California 94720} 
\affiliation{University of Liverpool, Liverpool L69 7ZE, United Kingdom} 
\affiliation{University College London, London WC1E 6BT, United Kingdom} 
\affiliation{Centro de Investigaciones Energeticas Medioambientales y Tecnologicas, E-28040 Madrid, Spain} 
\affiliation{Massachusetts Institute of Technology, Cambridge, Massachusetts  02139} 
\affiliation{Institute of Particle Physics: McGill University, Montr\'{e}al, Canada H3A~2T8; and University of Toronto, Toronto, Canada M5S~1A7} 
\affiliation{University of Michigan, Ann Arbor, Michigan 48109} 
\affiliation{Michigan State University, East Lansing, Michigan  48824} 
\affiliation{University of New Mexico, Albuquerque, New Mexico 87131} 
\affiliation{Northwestern University, Evanston, Illinois  60208} 
\affiliation{The Ohio State University, Columbus, Ohio  43210} 
\affiliation{Okayama University, Okayama 700-8530, Japan} 
\affiliation{Osaka City University, Osaka 588, Japan} 
\affiliation{University of Oxford, Oxford OX1 3RH, United Kingdom} 
\affiliation{University of Padova, Istituto Nazionale di Fisica Nucleare, Sezione di Padova-Trento, I-35131 Padova, Italy} 
\affiliation{LPNHE, Universite Pierre et Marie Curie/IN2P3-CNRS, UMR7585, Paris, F-75252 France} 
\affiliation{University of Pennsylvania, Philadelphia, Pennsylvania 19104} 
\affiliation{Istituto Nazionale di Fisica Nucleare Pisa, Universities of Pisa, Siena and Scuola Normale Superiore, I-56127 Pisa, Italy} 
\affiliation{University of Pittsburgh, Pittsburgh, Pennsylvania 15260} 
\affiliation{Purdue University, West Lafayette, Indiana 47907} 
\affiliation{University of Rochester, Rochester, New York 14627} 
\affiliation{The Rockefeller University, New York, New York 10021} 
\affiliation{Istituto Nazionale di Fisica Nucleare, Sezione di Roma 1, University of Rome ``La Sapienza," I-00185 Roma, Italy} 
\affiliation{Rutgers University, Piscataway, New Jersey 08855} 
\affiliation{Texas A\&M University, College Station, Texas 77843} 
\affiliation{Istituto Nazionale di Fisica Nucleare, University of Trieste/\ Udine, Italy} 
\affiliation{University of Tsukuba, Tsukuba, Ibaraki 305, Japan} 
\affiliation{Tufts University, Medford, Massachusetts 02155} 
\affiliation{Waseda University, Tokyo 169, Japan} 
\affiliation{Wayne State University, Detroit, Michigan  48201} 
\affiliation{University of Wisconsin, Madison, Wisconsin 53706} 
\affiliation{Yale University, New Haven, Connecticut 06520} 
\author{T.~Aaltonen}
\affiliation{Division of High Energy Physics, Department of Physics, University of Helsinki and Helsinki Institute of Physics, FIN-00014, Helsinki, Finland}
\author{A.~Abulencia}
\affiliation{University of Illinois, Urbana, Illinois 61801}
\author{J.~Adelman}
\affiliation{Enrico Fermi Institute, University of Chicago, Chicago, Illinois 60637}
\author{T.~Akimoto}
\affiliation{University of Tsukuba, Tsukuba, Ibaraki 305, Japan}
\author{M.G.~Albrow}
\affiliation{Fermi National Accelerator Laboratory, Batavia, Illinois 60510}
\author{B.~\'{A}lvarez~Gonz\'{a}lez}
\affiliation{Instituto de Fisica de Cantabria, CSIC-University of Cantabria, 39005 Santander, Spain}
\author{S.~Amerio}
\affiliation{University of Padova, Istituto Nazionale di Fisica Nucleare, Sezione di Padova-Trento, I-35131 Padova, Italy}
\author{D.~Amidei}
\affiliation{University of Michigan, Ann Arbor, Michigan 48109}
\author{A.~Anastassov}
\affiliation{Rutgers University, Piscataway, New Jersey 08855}
\author{A.~Annovi}
\affiliation{Laboratori Nazionali di Frascati, Istituto Nazionale di Fisica Nucleare, I-00044 Frascati, Italy}
% ...
\collaboration{CDF Collaboration\footnote{With visitors from $^a$University of Athens, 15784 Athens, Greece, 
$^b$Chinese Academy of Sciences, Beijing 100864, China, 
$^c$University of Bristol, Bristol BS8 1TL, United Kingdom, 
$^d$University Libre de Bruxelles, B-1050 Brussels, Belgium, 
$^e$University of California Irvine, Irvine, CA  92697, 
$^f$University of California Santa Cruz, Santa Cruz, CA  95064, 
$^g$Cornell University, Ithaca, NY  14853, 
$^h$University of Cyprus, Nicosia CY-1678, Cyprus, 
$^i$University College Dublin, Dublin 4, Ireland, 
$^j$University of Edinburgh, Edinburgh EH9 3JZ, United Kingdom, 
$^k$University of Heidelberg, D-69120 Heidelberg, Germany, 
$^l$Universidad Iberoamericana, Mexico D.F., Mexico, 
$^m$University of Manchester, Manchester M13 9PL, England, 
$^n$Nagasaki Institute of Applied Science, Nagasaki, Japan, 
$^o$University de Oviedo, E-33007 Oviedo, Spain, 
$^p$Queen Mary, University of London, London, E1 4NS, England, 
$^q$Texas Tech University, Lubbock, TX  79409, 
$^r$IFIC(CSIC-Universitat de Valencia), 46071 Valencia, Spain, 
}}
\noaffiliation
\pacs{14.40.Nd, 12.40.Yx}

\begin{abstract}
We report the observation of two narrow resonances
consistent with states of orbitally excited ($L=1$)
$B_s$ mesons using $1$ $\mathrm{fb^{-1}}$ of $p\overline{p}$ 
collisions at $\sqrt{s}$ = 1.96 TeV collected
with the CDF II detector at the Fermilab Tevatron.
We use two-body decays into $K^-$ and $B^+$ mesons 
reconstructed as $B^+\rightarrow J/\psi\,K^+$,
$J/\psi\rightarrow\mu^+\mu^-$ or $B^+\rightarrow \overline{D}^0\pi^+$,
$\overline{D}^0\rightarrow K^+\pi^-$.
We deduce the masses of the two states to be
$m(B_{s1}) = 5829.4 \pm  0.7$ $\mathrm{MeV}/c^2$ and
$m(B_{s2}^*) = 5839.6 \pm 0.7$ $\mathrm{MeV}/c^2$.
\end{abstract}

\maketitle

The heavy mesons consisting of a light and a heavy quark form
an interesting laboratory for the study of
QCD, 
the theory of the strong interaction. 
They are a close analogue to the hydrogen atom and
play a similar role for the study of the QCD as hydrogen 
for quantum electrodynamics.
According to heavy quark effective
theory (HQET) \cite{HQET},
in the limit of infinite mass of the heavy quark,
the heavy quark decouples from the degrees of freedom
of the light quark.
For orbitally excited states ($L=1$), the total
angular momentum of the light quark is $j_q=1/2$ or $j_q=3/2$.
Combining $j_q$ with the spin of the heavy quark, four states forming
two $j_q$ doublets are expected. 
For an infinite mass of heavy quark, the four states are 
degenerate in mass.
However, corrections due to finite mass 
lead to fine structure splitting between the two doublets 
and hyperfine structure
splitting within each of the doublets.
Following the standard scheme \cite{PDBook}, the states
with $j_q=1/2$ are named $B_{s0}^*$ and $B_{s1}$, and
the states with $j_q=3/2$ are named $B_{s1}$ and $B_{s2}^*$.
Often these four states are
referred to as $B_{s}^{**}$.

If kinematically allowed, all four $B_{s}^{**}$ states are expected
to decay dominantly to $B^{*}K$, $BK$ or both.
The states of the doublet with $j_q=1/2$
decay through an S-wave transition and are therefore expected to
have broad mass distributions.
The states with $j_q=3/2$ decay through a D-wave
transition and therefore are expected to 
have narrow mass distributions. 
In the following we focus on the narrow doublet.
While the $B_{s1}$ decays only into $B^*K$ due to 
conservation of spin and  parity, the $B_{s2}^{*}$
can decay to $B^{*}K$ and $BK$. 
If the $B_{s2}^{*}$ mass is
near the $B^{*}K$ threshold, the decay to  $B^{*}K$ will be
strongly suppressed compared to the decay to $BK$ due to the
available phase space.

Several theoretical predictions for the basic properties of the $B_{s}^{**}$ states
are available \cite{theory1}.
The predictions for the $B_{s1}$  mass range
from $5805$ to $5891$ $\mathrm{MeV}/c^2$, and for $B_{s2}^*$ $5820$ to $5903$ $\mathrm{MeV}/c^2$
with a mass difference between both states in the range $12$ to $20$ $\mathrm{MeV}/c^2$.
The natural widths of the two states are
expected to be of order $1$ $\mathrm{MeV}/c^2$ with
strong variation with predicted mass.

While there is already considerable information about $D_{s}^{**}$ mesons,
the analogous particles in the charm sector
\cite{PDBook, Bfactories1}, 
experimental knowledge about the $B_{s}^{**}$ mesons is minimal.
First evidence for at least one of the $B_s^{**}$ states was found by the OPAL experiment
\cite{Opal}. Evidence for a single state interpreted as $B_{s2}^*$ was seen by the Delphi 
Collaboration \cite{Delphi} and a preliminary observation of this state was reported recently 
by the D\O{} Collaboration \cite{D0}.

In this Letter, we report on the
observation of two states consistent with the $j_q=3/2$ doublet of the
$B_s^{**}$ decaying to $B^{+}K^-$ and $B^{*+}K^-$ 
with $B^{*+}\rightarrow B^+\gamma$, where the photon is not detected.
Because of the missing photon, the observed $B_{s1}$ peak is
shifted downward in mass by the $B^{*}$-$B$ mass splitting of 
$45.78 \pm 0.35$ $\mathrm{MeV}/c^2$ \cite{PDBook}.
$B^+$ mesons are reconstructed in two decay channels,
$B^+\rightarrow J/\psi \, K^+$ with $J/\psi\rightarrow
\mu^+\mu^-$ and $B^+\rightarrow\overline{D}^0\pi^+$ with
$\overline{D}^0\rightarrow K^+\pi^-$.
The use of a specific particle state implies the use of the 
charge-conjugate state as well. 
We use data collected by the CDF II Detector at
the Fermilab Tevatron between February 2002 and February 2006
corresponding to a total integrated luminosity of
$1$ $\mathrm{fb^{-1}}$.

The components of the CDF II detector \cite{Acosta:2004yw} 
used for this analysis are the magnetic spectrometer and
the muon detectors. 
The tracking system
is composed of a silicon microstrip detector
\cite{Hill:2004qb}
surrounded
by an open-cell drift chamber (COT) \cite{Affolder:2003ep}.
Both components are located inside a $1.4$ T
axial magnetic field. Muons are detected in planes of
multi-wire drift chambers and scintillators
\cite{Ascoli:1987av} in the pseudorapidity range $\mathrm{|\eta|
\le 1.0}$, where $\eta=-\ln\tan(\theta/2)$ and $\theta$ is
the polar angle measured from the proton beam direction.
Hadron identification is crucial for distinguishing kaons 
originating from $B_{s}^{**}$ decays from other particles.
It is provided by a combination of the ionization energy loss 
in the COT and a measurement by a time-of-flight
system \cite{TOF}.

A three-level trigger system is used for the online event
selection.  
The level 1 trigger system includes the eXtremely Fast
Tracker (XFT) track processor
\cite{XFT} which finds charged-particle tracks in the COT
and measures their azimuthal angle around the beam direction 
and transverse momenta.  In 
level 2, the silicon vertex trigger \cite{SVT} adds hits from the silicon
detector to tracks found by the XFT to 
provide measurements of impact parameter.
The level 3 system confirms the selections using a version of the offline
event reconstruction optimized for speed.

The dimuon trigger \cite{Acosta:2004yw} requires two tracks of opposite charge matched to track segments in the muon chambers, where the mass of the pair is consistent with the $J/\psi$ mass.
The displaced-vertex trigger \cite{DTrig} requires two tracks
with large impact parameters. Additionally, the intersection of the tracks has to be displaced from the interaction point and a minimum transverse momentum, the momentum component perpendicular to 
the proton beam direction, is required for each track.


In both samples, we
reconstruct $B_s^{**}$ candidates by combining
$B^+$ candidates with $K^-$ candidates.
In the dimuon (displaced-vertex) trigger sample, we form
$J/\psi\rightarrow \mu^+\mu^-$ ($\overline{D}^0\rightarrow K^+\pi^-$) candidates 
and combine each $J/\psi$ ($\overline{D}^0$) candidate
with a track assumed to be a kaon (pion), constraining the tracks to 
an appropriate decay topology
to form a $B^+$ candidate.
At this stage hadron identification is 
not used.
In order to improve the mass resolution, we
consider the quantity $Q$ defined as
\mbox{$m(B^+K^-)-m(B^+)-M_{K^-}$}, where $m(B^+K^-)$ and $m(B^+)$ are
the reconstructed invariant
masses of the $B^+\,K^-$ pair and the $B^+$ candidate, and 
$M_{K^-}$ is the known kaon mass \cite{PDBook}.
The predicted $B_{s1}$ ($B_{s2}^*$) state mass translates 
to the region $0 < Q < 73$
$\mathrm{MeV}/c^2$ ($48\,<\,Q\,<\,131$ $\mathrm{MeV}/c^2$).

For the selection of candidates, we use a chain of two neural
networks based on the {\sc NeuroBayes}
\cite{NeuroB} package in each of the $B^+$ decay channels. 
In a first step, a neural network in each channel combines
topological, kinematic, and particle identification
quantities for the $B^+$ and its daughters
to form a single discriminant between $B^+$ mesons
and background. 
The most important quantities are
the impact parameter of the $B^{+}$,
the projection of the displacement of its reconstructed
decay point from the beamline
on the direction of its transverse momentum,
the transverse momentum of the $B^+$ decay's pion (kaon),
and its impact parameter.
The neural networks are trained on two classes of events corresponding
to the signal and background samples.
In the $B^{+} \rightarrow J/\psi\,K^{+}$ channel, we use a
{\sc pythia} \cite{Pythia} simulation for the signal sample and
experimental data from the $B^+$ mass sidebands
$5190$ - $5240$ $\mathrm{MeV}/c^2$ and  
$5320$ - $5395$ $\mathrm{MeV}/c^2$ for the background sample.
In the $B^{+} \rightarrow \overline{D}^0 \pi^{+}$ channel
we use only experimental data to train the
$B^+$ neural network. We use candidates from a signal region between
$5240$ and $5310$ $\mathrm{MeV}/c^2$
in the invariant mass as signal sample and data from a
$B^+$ mass sideband between
$5325$ and $5370$ $\mathrm{MeV}/c^2$
as background sample. 
The events from the $B^+$ mass sidebands are used also
as signal with negative weight to account for
the background in the signal region. 
Based on the neural networks, we select approximately $31000$
$B^+$ signal events in the $J/\psi\, K^+$ decay
channel and $27200$ in the $\overline{D}^0\pi^+$ channel.

\begin{figure}[htb]
\begin{center}
 \includegraphics[width=8.0cm]{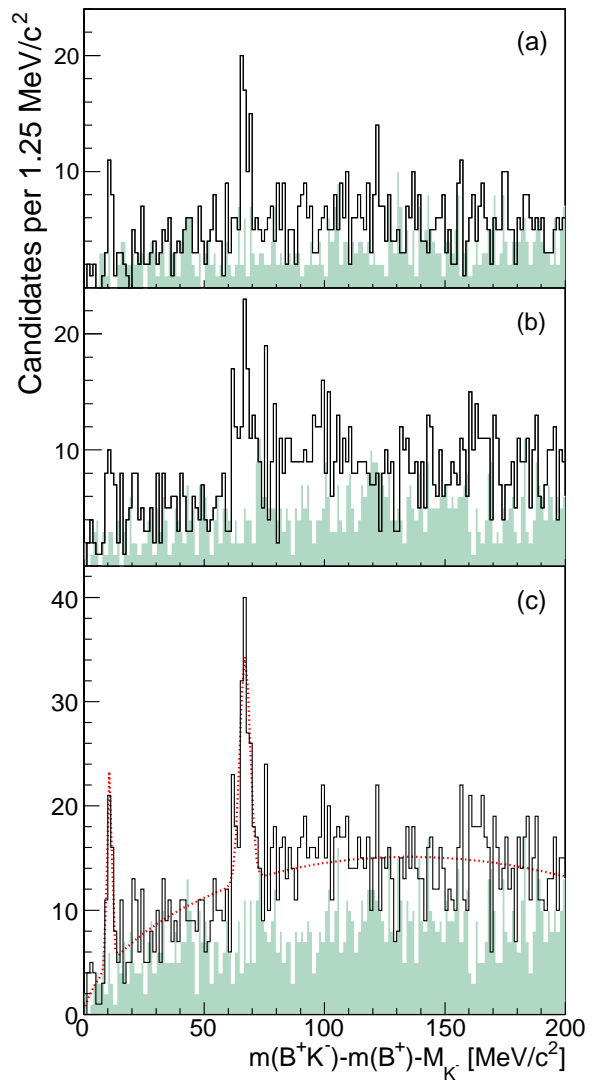}
 \caption{Distribution of 
\mbox{$Q=m(B^+K^-)-m(B^+)-M_{K^-}$}
for the $B_s^{**}$ candidates with (a) $B^+\rightarrow J/\psi\,
K^+$, (b) $B^+\rightarrow \overline{D}^0\pi^+$ and (c) both
$B^+$ channels combined. The dotted line shows the result of a
fit with the sum of a background function and two Gaussians.
The filled area shows the $Q$ distribution for the
wrong-sign combination $B^{+} K^{+}$.\label{fig2}}
\end{center}
\end{figure}

In a second step, we
select $B^{**}_{s}$ candidates based on the number of candidates per event and 
on an additional neural network for each $B^+$ decay channel.
These neural networks use
the same inputs as used by the
neural networks to select $B^+$ mesons as well as their discriminant, 
and kinematic and particle identification quantities for
the kaon track of the $B_{s}^{**}$ decay. 
The particle identification of the kaon is the most important variable, 
followed by the neural network discriminant of the $B^+$ and the pseudorapidity of the
kaon.
The number of candidates per event is
not used in the neural network due to the difficulty of
modeling fragmentation and hadronization in the production
of heavy quarks.  We select
only those events with fewer than four candidates 
because a lower number of candidates
provides  a better signal-to-background ratio.
This cut is fixed without looking to the experimental data, based only upon the
above assumption.
The $B_s^{**}$ neural networks are trained on a combination of
simulated events, containing only signal, 
and data events in the $Q$ range $0$ to $200$
$\mathrm{MeV}/c^2$ for background sample. 
The number of real $B_{s}^{**}$ mesons in the background-training
sample is too small to affect significantly the learning process of the neural network.
In order to avoid possible mass biases,
the simulated signal events
have the same $B^{+} K^{-}$ mass distribution
as the events used for background in the neural network training.
The value of the cut on the neural network
discriminant for the final selection is chosen to optimize
$N_{MC}/\sqrt{N_{data}}$, where $N_{data}$ ($N_{MC}$) is the number of the
selected candidates in data (simulation) in the $Q$
range $60$ to $70$ $\mathrm{MeV}/c^2$. 
This range has been chosen based on the mass the of previously seen
$B_{s2}^*$ state and therefore
is not biased with respect to the unobserved state. We verify that the
observed $B_s^{**}$ masses do not depend on the $Q$ range
used for cut optimization.
The $Q$ distributions of the selected candidates are shown in 
Figs.~\ref{fig2}(a) and \ref{fig2}(b) for the two trigger samples separately,
and in Fig.~\ref{fig2}(c) added together. 
Two peaks are visible, centered near $67$ $\mathrm{MeV}/c^2$ 
and near $10$ $\mathrm{MeV}/c^2$. 
The wrong-sign combinations (filled area in Fig. \ref{fig2}) do not show any significant structure.

The two peaks in data can be interpreted as the two $j_q = 3/2$
states of orbitally excited $B_s$
mesons. The natural interpretation is that the peak near
$67$ $\mathrm{MeV}/c^2$ stems from the 
$B_{s2}^*\rightarrow B^+ K^-$ decay while the peak near
$10$ $\mathrm{MeV}/c^2$  stems from the decay
$B_{s1}\rightarrow B^{*+}K^-$. 
Reversing the assignment of the two peaks
would result in a larger mass difference between 
$B_{s2}^*$ and $B_{s1}$ with $B_{s1}$ being heavier, 
which would be opposite to other heavy quark mesons.

To extract the mean $Q$ values for the two peaks, we use an unbinned maximum
likelihood fit. Each of the peaks is described by a
Gaussian shape.
We use a phenomenological function to describe the background
without distinguishing different types of
backgrounds.
The functional form of the background shape is
$[Q(\beta -Q)]^\gamma
\exp{(-\gamma Q)}$, 
where $\beta$ and $\gamma$ are free
parameters.
The fit has three free parameters for each of the Gaussians and
two free parameters for the background. 
The fit to each data sample separately gives results consistent 
between the two $B^+$ decay channels.
Therefore, we combine the $B^+$ channels to perform the final fit.
The projection of the fit on the full sample is shown in 
Fig.~\ref{fig2}(c). From the fit we extract $Q(B_{s1}) = 10.73
\pm 0.21$ $\mathrm{MeV}/c^2$ and
$Q(B_{s2}^*) = 66.96 \pm 
0.39$ $\mathrm{MeV}/c^2$,
with yields $N(B_{s1}) = 36 \pm 9$ events and
$N(B_{s2}^*) = 95 \pm 23$ events, where all uncertainties are statistical.
The widths of the Gaussians are consistent with
expected detector resolutions. 

Systematic uncertainties on the $Q$ value may arise 
from the tracking and fitting procedures.
The sources of uncertainty from the tracking are
the uncertainty on the track error matrix, which enters
through vertex fits, and the uncertainty on the material and
magnetic field distribution inside the tracking volume.
Based on a detailed study performed for 
the measurement of mass and
width of the orbitally excited $D^{**}$ states
\cite{CDFDSS}, we assign a combined systematic uncertainty
of $0.14$ $\mathrm{MeV}/c^2$ due to the tracking effects.
We study effects on the fitting procedure of
the unknown background shape
and the simplification by the single-Gaussian signal description.
In both cases we generate a large number of samples of the
same size as the data. For the background shape study we use a
probability density function proportional
to a 
fit on data with a third order polynomial as background function
instead of the default one for sample generation.
In the signal shape study the sum of two Gaussians,
which have the width one gets by fitting each 
decay channel separately, is used for sample generation.
Each of the samples is then
fitted using the default fit model and the pull distributions
are examined. In both cases the pull distributions are
consistent with a Gaussian with a mean of zero and unit width.
Therefore, we do not assign any systematic
uncertainty arising from the fitting procedure.
The resulting $Q$ values are
\begin{eqnarray}
Q(B_{s1})\,&=&\,10.73\,\pm\,0.21\,\mathrm{(stat)}\,\pm\,0.14\,\mathrm{(sys)}\,\mathrm{MeV}/c^2
\nonumber ,\\
Q(B_{s2}^*)\,&=&\,66.96\,\pm\,0.39\,\mathrm{(stat)}\,\pm\,0.14\,\mathrm{(sys)}\,\mathrm{MeV}/c^2
\nonumber .
\end{eqnarray}
By adding the known values \cite{PDBook} of $M_{B^{*}}$ and
$M_{K^-}$ to
$Q(B_{s1})$ and
$M_{B^+}$ with
$M_{K^-}$ to $Q(B_{s2}^*)$,
we obtain
$m(B_{s1}) = 5829.4 \pm 0.7$ $\mathrm{MeV}/c^2$
and 
$m(B_{s2}^*) = 5839.6 \pm 0.7$ $\mathrm{MeV}/c^2$.
The statistical and systematic uncertainties on the $Q$ value
and the uncertainties on the masses of $K^-$ and $B^{+}$ or $B^{*+}$
are added in quadrature.
Finally, the mass difference of the
two narrow $B_s^{**}$ states is
$\Delta
m(B_{s2}^*,B_{s1}) = 10.5 \pm 0.6$ $\mathrm{MeV}/c^2$, 
where we add $M_{B^{*}} - M_{B} = 45.78 \pm 0.35$
$\mathrm{MeV}/c^2$ to the difference of the two measured $Q$
values.

To estimate the statistical significance of each of the two peaks, we
repeat the fit without the term for one of the Gaussian peaks
and again without the other peak.
For each peak we form $\Delta\mathcal{L}=-2\ln{\mathcal{L}_0/\mathcal{L}}$,
where $\mathcal{L}$ is the value of the likelihood function of the original fit
and $\mathcal{L}_0$ is the value for the fit without one of the peaks,
and measure
$\Delta\mathcal{L}=48.7$ ($74.5$) for $B_{s1}$
($B_{s2}^*$). 
We generate samples with
background according to the background function from the fit to
the data and one of the peaks.
For each of the samples, we
evaluate $\Delta\mathcal{L}$ . 
In the $\Delta\mathcal{L}$ distribution for 
samples created without $B_{s1}$ ($B_{s2}^*$)
the highest observed value is 35.2 (41.8)
in over 4.6 (5.6) million samples
where the peak is located in range
$0$ - $50$ $\mathrm{MeV}/c^2$ ($20$ - $120$ $\mathrm{MeV}/c^2$).
Therefore, we conclude that the statistical significance of each
of the signals exceeds five standard deviations.

In summary, we report the first observation of
the narrow $j_q=3/2$
states of the orbitally excited $B_s$ mesons. The  
signals observed are attributed to the $B_{s2}^*\rightarrow B^+K^-$
and $B_{s1}\rightarrow B^{*+}K^-$ decays. 
From the precise measurement of the $Q$ values, we derive the
masses of the two states and their mass difference, and the
values are consistent with theoretical predictions \cite{theory1}.

We thank the Fermilab staff and the technical staffs of the participating 
institutions for their vital contributions. This work was supported by 
the U.S. Department of Energy and National Science Foundation; the 
Italian Istituto Nazionale di Fisica Nucleare; the Ministry of Education, 
Culture, Sports, Science and Technology of Japan; the Natural Sciences 
and Engineering Research Council of Canada; the National Science Council 
of the Republic of China; the Swiss National Science Foundation; the A.P. 
Sloan Foundation; the Bundesministerium f\"ur Bildung und Forschung, 
Germany; the Korean Science and Engineering Foundation and the Korean 
Research Foundation; the Science and Technology Facilities Council and the Royal 
Society, UK; the Institut National de Physique Nucleaire et 
Physique des Particules/CNRS; the Russian Foundation for Basic Research; 
the Comisi\'on Interministerial de Ciencia y Tecnolog\'{\i}a, Spain;
the European Community's Human Potential Programme under
contract HPRN-CT-2002-00292; the Slovak R\&D Agency; and
the Academy of Finland.


\end{document}